\documentclass[aps,prd,twocolumn,a4paper,superscriptaddress]{revtex4-1}
\usepackage{graphicx}
\begin{document}

\newcommand{\be}{\begin{equation}}
\newcommand{\ee}{\end{equation}}
\newcommand{\bq}{\begin{eqnarray}}
\newcommand{\eq}{\end{eqnarray}}
\newcommand{\bsq}{\begin{subequations}}
\newcommand{\esq}{\end{subequations}}
\newcommand{\bc}{\begin{center}}
\newcommand{\ec}{\end{center}}

\title{Physical and Invariant Models for Defect Network Evolution}

\author{C. J. A. P. Martins}
\email{Carlos.Martins@astro.up.pt}
\affiliation{Centro de Astrof\'{\i}sica da Universidade do Porto, Rua das Estrelas, 4150-762 Porto, Portugal}
\affiliation{Instituto de Astrof\'{\i}sica e Ci\^encias do Espa\c co, CAUP, Rua das Estrelas, 4150-762 Porto, Portugal}
\author{M. M. P. V. P. Cabral}
\email[]{up201203567@fc.up.pt}
\affiliation{Centro de Astrof\'{\i}sica da Universidade do Porto, Rua das Estrelas, 4150-762 Porto, Portugal}
\affiliation{Faculdade de Ci\^encias, Universidade do Porto, Rua do Campo Alegre 687, 4169-007 Porto, Portugal}

\date{14 January 2016}

\begin{abstract}

We revisit the velocity-dependent one-scale model for topological defect evolution, and present a new alternative formulation in terms of a physical (rather than invariant) characteristic length scale. While the two approaches are equivalent (as we explicitly demonstrate), the new one is particularly relevant when studying the evolution of ultra-relativistic defects. Moreover, a comparison of the two provides further insight on the interpretation of the model's two phenomenological parameters, $c$ related to energy losses and $k$ related to the curvature of the defects. As an illustration of the relevance of the new formulation, we use it to study the evolution of cosmic string and domain wall networks in contracting universes. We show that these networks are ultra-relativistic and conformally contracted, with the physical length scale behaving as $L_{ph}\propto a$ and the density as $\rho\propto a^{-4}$ (as in a radiation fluid) in both cases. On the other hand the velocity and invariant length respectively behave as $(\gamma v)\propto a^{-n}$ and $L_{inv}\propto a^{\frac{4}{4-n}}$, where $n$ is the number of dimensions of the defect's worldsheet. Finally we also study an alternative friction-dominated scenario and show that the stretching and Kibble regimes identified in the case of expanding universes can also occur for contracting ones.

\end{abstract}
\maketitle

\section{Introduction}

It is thought that the early universe underwent a series of phase transitions, each one spontaneously breaking some symmetry in particle physics and giving rise to topological  defects of some kind \cite{Kibble1,Book}. In many cases the defects will persist throughout the subsequent evolution of the universe, providing fossil relics of its early stages. Understanding the evolution and cosmological consequences of these defect networks is therefore a mandatory component of any consistent attempt to quantitatively describe the early universe.

Since defects are intrinsically non-linear objects, there are two basic approaches to studying their evolution. The first is to resort to field theory (and/or, for cosmic strings, Goto-Nambu) numerical simulations \cite{PRS,BENBOU,ALBTUR,ALESHE,Mcgraw98,ABELIAN,Yamaguchi02,MS3,Garagounis03,MS4,Ringeval07,Olum07,Urrestilla08,Stuckey09,Pillado11,Leite12,Hiramatsu13,Achucarro14}. Alternatively, one can develop analytic models which aim to capture the key properties of the network, at least on large scales.

The first effort within the latter framework was Kibble's one-scale model of string networks \cite{KIB}, which has a single macroscopic parameter: a length scale that can be identified as the string correlation length, the string curvature radius, or the inter-string distance---in the model's approximation they should be seen as identical or at least comparable. This was later generalized to a three-scale model by Austin, Copeland and Kibble \cite{ACK} where there are three distinct length scales: the first two are again the string correlation length and the inter-string distance, while the third is a typical scale for small-scale structures.

A different approach stems from the realization that in order to be able to quantitatively describe the whole cosmological history of these networks one must be able to describe the evolution of the defect velocities, not least because depending on the cosmological epoch (and on their own specific properties) the defects may be moving at non-relativistic or at ultra-relativistic speeds. This is the basis for the velocity-dependent one-scale model (VOS) of Martins and Shellard \cite{MS1,MS2}, which retains Kibble's assumptions on the existence of a single length scale but adds the RMS velocity as a second macroscopic quantity. This model has been rigorously derived starting from the Goto-Nambu action, and it has also been successfully tested and calibrated against high-resolution field theory \cite{ABELIAN,MS3} and Goto-Nambu numerical simulations \cite{MS3,MS4}. More recently, it has also been extended to domain walls \cite{AWALL}, monopoles \cite{MONOP}, semilocal strings \cite{SEMIL} and several other related contexts. 

The standard VOS model relies on an invariant length scale (to be rigorously defined below), although in a few specific circumstances a corresponding physical length scale was also briefly discussed---in particular, in \cite{Cyclic,Contracting}. Here we present a detailed formulation of the VOS model in terms of a physical characteristic length scale. We also show that the two descriptions are equivalent (leading to the same scaling solutions), and discuss how the model parameters are related in the two approaches. A model based on a physical length scale turns out to be particularly useful for studying the evolution of ultra-relativistic defects. As a concrete illustration of this point, we revisit the evolution of cosmic strings in contracting universes, first studied in \cite{Cyclic,Contracting} (whose results we generalize) and further extend it by discussing case for domain wall networks.

\section{Analytic Modeling Generalities}

Broadly speaking, the key idea behind the analytic modeling of defect networks is that one abandons the idea of studying their `statistical physics' (which can only be done numerically, except in idealized circumstances) and instead concentrates on its `thermodynamics'. In other works, one or more macroscopic quantities are chosen to describe the network, and the knowledge of the microphysics is used to obtain the evolution equations for these macroscopic quantities. This has the advantage of leading to relatively simple models which can, in principle, encapsulate most of the relevant physics, but it also has an associated cost: in going from the microphysics to the macrophysics one is forced to introduce phenomenological parameters, which can only be calibrated by direct comparison with numerical simulations. In what follows we provide a simplified derivation of the dynamical equations of the model, along the lines of \cite{MONOP}.

Consider a network of defects with $n$-dimensional worldsheets ($n=1$ for monopoles, $n=2$ for cosmic strings and $n=3$ for domain walls) evolving in $(3+1)$ space-time dimensions.  We temporarily assume them to have velocity $v$ and to be non-interacting and (for the case of extended objects) planar. Then the momentum per unit comoving defect volume---simply the momentum, in the case of monopoles---behaves as
\be
p\propto a^{-1}\Longrightarrow v\gamma\propto a^{-n}
\ee
from which we get, by differentiation
\be
\frac{dv}{dt}+nH(1-v^2)v=0\,.
\ee
Under the above hypotheses the average number of defects in a fixed comoving volume should be conserved, which implies
\be
\rho\propto \gamma a^{-(4-n)}
\ee
and again, differentiating and using the velocity equation, we find
\be
\frac{d\rho}{dt}+H[(4-n)+nv^2]\rho=0\,.
\ee
The hypotheses so far are, of course, widely unrealistic. However, we can use this as a starting point to build a reasonable model. As has been pointed out above, the validity of this process can be checked for the case of cosmic strings, where a more rigorous derivation has been done \cite{MS1,MS2}.

Let us  start by defining a characteristic length scale
\be
L^{4-n}=\frac{M}{\rho}\,, \label{charscale}
\ee
where $M$ will have dimensions appropriate for the defect in question (i.e., monopole mass, string mass per unit length, or wall mass per unit area), and can also be written
\be
M\sim \eta^n, \label{massscale}
\ee
with $\eta$ being the corresponding symmetry breaking scale. Also, we interpret the velocity as being the RMS velocity of the defect network, and allow for energy losses due to interactions, which in many physically relevant circumstances can be modeled (purely on dimensional grounds) by
\be
\frac{d\rho}{dt}=-c\frac{v}{L}\rho\,. \label{energyloss}
\ee
An additional source of defect damping is friction due to particle scattering. Typically this is only relevant in the early stages of cosmological defect evolution, but we introduce it here for completeness. This damping can be characterized by a friction length scale
\be
{\bf F}=-\frac{M}{\ell_f}\gamma {\bf v}\,, \label{fricscale}
\ee
where we are defining
\be\label{deftheta}
\ell_f\equiv \frac{M}{\theta T^{n+1}}\propto a^{n+1}\,,
\ee
$T$ is the background temperature and $\theta$ is a parameter counting the number of particle degrees of freedom interacting with the defects. We can also define an overall damping length which includes both the effect of Hubble damping and the friction due to particle scattering
\be
\frac{1}{\ell_d}=nH+\frac{1}{\ell_f}\,.
\ee

Putting together all of the above effects, we find the following evolution equations for the characteristic length scale $L$ and RMS velocity $v$
\be
(4-n)\frac{dL}{dt}=(4-n)HL+v^2\frac{L}{\ell_d}+cv
\ee
\be
\frac{dv}{dt}=(1-v^2)\left[f-\frac{v}{\ell_d}\right]
\ee
where in the latter equation we have also allowed for the possibility of further driving forces affecting the defect dynamics. Note that $f$ has the units of acceleration---it is the force per unit mass. For extended objects (walls and strings) that have been extensively studied in the past, this driving force is obviously the local curvature, and we have
\be
f\sim\frac{k}{R}=\frac{k}{L}\,;
\ee
in the last equality we are implicitly assuming that our characteristic length scale is the same as the defect curvature radius. For monopoles the situation is somewhat more complex since there are forces due to other monopoles, as discussed in \cite{MONOP}.

\section{Physical and invariant quantities}

Throughout the previous section, the characteristic length scale $L$ was an invariant quantity or, in other words, a measure of the invariant string energy (and hence length). We now discuss how to express the VOS model in terms of a physical length scale. Let us start by considering the network's energy. The invariant and physical quantities are related through the standard Lorentz factor, $\gamma=(1-v^2)^{-1/2}$, as follows
\be
E_{inv}=\gamma E_{ph}\,. 
\ee
Since, according to Eq. (\ref{charscale}), $\rho\propto L^{-(4-n)}$, we see that the characteristic length scales are related via
\be
L_{ph}=\gamma^{\frac{1}{4-n}}L_{inv}\,. \label{relphinv}
\ee

Note that this length scale is a measure of the total energy content of the network, or (in the context of the VOS model assumption of a single independent characteristic scale) the typical separation between defects. We may instead define a characteristic defect size
\be
S_{inv}=\gamma S_{ph}\,;
\ee
this would therefore be a a characteristic length (or total length) for the strings, and a characteristic area (or total area) for walls. This can then be equivalently expressed in terms of a characteristic radius $S\propto R^{n-1}$, leading to
\be
R_{inv}=\gamma^{\frac{1}{n-1}}R_{ph}\,.
\ee

Let us consider the standard VOS model described in the previous section. Again, the evolution equations are
\be
(4-n)\frac{dL_{inv}}{dt}=(4-n)HL_{inv}+v^2\frac{L_{inv}}{\ell_d} +cv \, \label{originalL}
\ee
\be
\frac{dv}{dt}=(1-v^2)\left[\frac{k}{R_{inv}}-\frac{v}{\ell_d}\right]\,.
\ee
For clarity, we have now explicitly identified the invariant quantities. As is well known \cite{MS1,MS2,AWALL}, for a universe whose scale factor grows as a power law, $a\propto t^\lambda$ (with $0<\lambda<1$), these equations have an attractor scaling solution 
\be
\left(\frac{L}{t}\right)^2\equiv\epsilon^2=\frac{k(k+c)}{n(4-n)\lambda(1-\lambda)}
\ee
\be
v^2=\frac{4-n}{n}\frac{1-\lambda}{\lambda}\frac{k}{k+c}\,.
\ee
Note that in this attractor solution frictional damping due to particle scattering is negligible compared to that due to the Hubble expansion. We can now change variables using
\be
\frac{d\gamma}{dt}=v\gamma^3\frac{dv}{dt}
\ee
leading to
\be
\frac{d(\gamma v)}{dt}=\frac{k\gamma}{R}-\frac{\gamma v}{\ell_d}
\ee
and
\be
(4-n)\frac{dL_{ph}}{dt}=(4-n)HL_{ph}+kv\frac{L_{ph}}{R_{inv}}+\gamma^{\frac{1}{4-n}}cv\,.
\ee
Finally, noting that in the canonical model $R$ is an invariant quantity which in a one-scale model context is identified as $R_{inv}\equiv L_{inv}$ and transforming it to the physical one, we finally obtain
\be
\frac{d(\gamma v)}{dt}=\frac{k\gamma^{1+\frac{1}{4-n}}}{L_{ph}}-\frac{\gamma v}{\ell_d}
\ee
\be
(4-n)\frac{dL_{ph}}{dt}=(4-n)HL_{ph}+v(k+c)\gamma^{\frac{1}{4-n}}\,.
\ee
Note that the damping length scale does not appear in the evolution equation for the physical length scale, but only in the one for the invariant length scale (as well as in the one for the velocity). If we now look for attractor scaling solutions we get
\be
\epsilon^2_{ph}=\gamma^{\frac{2}{4-n}}\epsilon^2_{inv}
\ee
\be
(\gamma v)^2_{ph}=\gamma^2 v^2_{inv}\,,
\ee
which is trivially correct and consistent given the various definitions above.

We finally need to confirm how the model parameters $c$ and $k$ behave as one switches between the physical and invariant approaches. Starting with the energy loss term $c$, one simply has to generalize the argument first made in \cite{KIB} and note that the probability $dP$ that a defect segment will encounter another segment in a time interval $dt$ should be given approximately by
\be
dP = - \frac{d \rho}{\rho}=(4-n) \frac{d L_{ph}}{L_{ph}} \sim c_{ph}\frac{v dt}{L_{ph}} \sim c_{ph} \frac{v dt}{\gamma^{\frac{1}{4-n}} L_{inv}}\,.
\label{probloop}
\ee
From this we infer that
\be
c_{ph}=\gamma^{\frac{1}{4-n}}c_{inv}\,.
\ee
For the specific case of strings, if (as suggested in \cite{Cyclic}) one has $c_{inv}=c_0\gamma^{-1/2}$ (with $c_0$ being a constant), then it follows that $c_{ph}=\gamma^{1/2}c_{inv}=c_0=const$. Our analysis shows that analogous results hold regardless of the defect dimensionality.

A similar argument can be made for the curvature parameter (whose phenomenology, in the specific case of strings, has been discussed in detail in \cite{ABELIAN,MS3,MS4}), leading to
\be
k_{ph}=\gamma^{\frac{1}{4-n}}k_{inv}\,;
\ee
though of course the physical interpretation of this parameter for defects other than strings is less clear, so the other cases should be treated by analogy. For the case of strings its behavior is phenomenologically well described by the following velocity dependence
\be
k(v)\equiv \frac{2\sqrt{2}}{\pi}(1-v^2)(1+2\sqrt{2}v^3)
\frac{1-8v^6}{1+8v^6}\,.
\label{kkvvv}
\ee
One important point pertaining to the behavior of this parameter is that $k\rightarrow0$ as $v\rightarrow1$, and we assume that the same holds for domain walls (for which no analogous expression is known).

With these relations between the physical and invariant model parameters, we can finally write
\be
(4-n)\frac{dL_{ph}}{dt}=(4-n)HL_{ph}+(c_{ph}+k_{ph})v\, \label{physL}
\ee
\be
\frac{dv}{dt}=(1-v^2)\left[\frac{k_{ph}}{L_{ph}}-\frac{v}{\ell_d}\right]\,,
\ee
or equivalently
\be
\frac{d(\gamma v)}{dt}=\frac{\gamma k_{ph}}{L_{ph}}-\frac{\gamma v}{\ell_d}\,,\label{physgv}
\ee
which are the evolution equations for the VOS model based on physical rather than invariant parameters.

\section{Defects in contracting universes}

As an application of the model, we now discuss the evolution of defect networks in contracting universes, clarifying and extending the results of \cite{Cyclic,Contracting}. Some possible motivations for studying contracting universe scenarios have been recently reviewed in \cite{Bouncing}. In this section we ignore the effects of friction due to particle scattering (in other words, assume $\theta=0$ and $\ell_f\rightarrow\infty$). The consequences of relaxing this assumption will be discussed in the following section.

The key physical difference between this case and the standard one is that in a contracting phase the Hubble parameter becomes negative---in other words it becomes an acceleration term (rather than a damping term). As a result the velocity will increase and the network will become ultra-relativistic, with $v\rightarrow1$. This is true even though in this limit we expect $k(v)\rightarrow0$, at least for extended objects (cosmic strings and domain walls).

In this case one easily finds from Eq. (\ref{physgv}) an asymptotic behavior
\be
\gamma v\propto a^{-n}\,,
\ee
or more simply
\be
\gamma \propto a^{-n}\,.
\ee
On the other hand, from Eq. (\ref{originalL}) one finds for the invariant length scale
\be
L_{inv}\propto a^{\frac{4}{4-n}}\,,
\ee
which can be re-expressed in terms of the corresponding physical scale
\be
L_{ph}=\gamma^{\frac{1}{4-n}}L_{inv}\propto a\,.
\ee
These asymptotic scaling laws were briefly described in \cite{addedA,addedB}; in what follows we will study in more detail the evolution of the networks as they approach this asymptotic state. This last relation, which can also be obtained directly from Eq. (\ref{physL}), agrees with the intuitive expectation that the defect network is being conformally contracted as the universe collapses. Similarly for the characteristic radius for extended defects we have 
\be
R_{ph}\propto a\,,
\ee
and therefore
\be
R_{inv}=\gamma^{\frac{1}{n-1}}R_{ph}\propto a^{-\frac{1}{n-1}}\,.
\ee
We also note that in all cases the network's energy density behaves as
\be
\rho\propto L_{inv}^{-(4-n)}\propto a^{-4}\,;
\ee
again this is to be expected: an ultra-relativistic network behaves as a radiation fluid. An interesting consequence of this is that, even if the defect network eventually dominates the energy density of the universe, the universe's contraction rate will still be radiation-like. In any case, as the temperature rises as approaches that of the defect-forming phase transition we expect the defects to effectively dissolve into the high-density background.

We can further quantify how this asymptotic scaling regime is approached. This corresponds to studying the behavior of Eqs. (\ref{physL}) and (\ref{physgv}) when the $c_{ph}$ and $k_{ph}$ terms provide a small but not entirely negligible contribution. In the former case, we can assume $L_{ph}=a f(a)$, where $f(a)$ is a correction factor, in which case Eq. (\ref{physL}) leads to
\be
(4-n)a\frac{df}{da}\frac{da}{dt}=c_{ph}\,.
\ee
Assuming for simplicity that in this regime the scale factor behaves as $a\propto (t_c-t)^\lambda$, where $t_c$ is the Big Crunch time and $0<\lambda<1$, this equation can be straightforwardly integrated, leading to
\be
L_{ph}\propto a\left[1-\frac{c_{ph}}{(1-\lambda)(4-n)}a^{\frac{1}{\lambda}-1}\right]\,,
\ee
and as expected the correction factor approaches unity as $a\rightarrow0$.

Similarly for Eq. (\ref{physgv}) we can assume that $\gamma v=a^{-n}g(a)$, which leads to
\be
\frac{dg}{da}\frac{da}{dt}=\frac{k_{ph}}{L_{ph}}g\,,
\ee
and making the same assumption on the behavior of the scale factor we find the approximate solution
\be
\gamma\propto a^{-n}\exp{\left[-\frac{k_{ph}}{1-\lambda} a^{\frac{1}{\lambda}-1}   \right]} \approx a^{-n} \left[1-\frac{k_{ph}}{1-\lambda}  a^{\frac{1}{\lambda}-1}  \right]\,.
\ee
Note that the form of the correction term is quite similar to that for the length scale equation, the two differences being that the $\gamma$ correction is independent of the defect dimensionality (there's no dependence on $n$) and that $k_{ph}$ is expected to be negative in this limit and to approach zero as $v\rightarrow1$ while $c_{ph}$ is expected to be a positive constant. Finally we can put the two together using Eq. (\ref{relphinv}), obtaining
\be
L_{inv}\propto a^{\frac{4}{4-n}} \left[1-\frac{c_{ph}-k_{ph}}{(1-\lambda)(4-n)}a^{\frac{1}{\lambda}-1}\right]\,,
\ee
which matches the physical intuition that $c_{ph}$ should be more important than $k_{ph}$ in determining this correction.

\section{Alternative scenario: friction domination}

In the previous section the effects of friction due to particle scattering were neglected. This is a reasonable assumption for heavy defects (say, those formed around  the GUT scale) since in that case friction is only significant very close to the defect-forming phase transition and becomes negligible soon afterwards \cite{MS1,MS2,AWALL}. Although this argument holds, in principle, both for strings and for domain walls, in practice it is only relevant for strings since heavy domain walls are observationally ruled out \cite{Zeldovich}. The same argument would naturally apply for a contracting universe. However, for very light defects (say, those formed around the electroweak scale) friction will dominate over Hubble damping for a considerable period. In what follows we discuss how the above solutions change in this case.

Let us start with some order of magnitude estimates. We are comparing the Hubble and friction contributions to the damping length scale
\be
\frac{1}{\ell_d}=nH+\frac{1}{\ell_f}=nH+\frac{\sigma}{a^{n+1}}\,,
\ee
where for convenience we wrote the friction length scale in terms of the scale factor by introducing a constant parameter $\sigma$ which is related to the parameter $\theta$ defined in Eq. \ref{deftheta}. Both of these parameters count the number of effective degrees of freedom which interact with the defect. To give an example, assuming that all standard model degrees freedom interact with a cosmic string formed at the GUT scale ($T\sim10^{16}$GeV), then at formation the friction term is 30 times larger than the Hubble term; for a string formed at the electroweak scale ($T\sim10^{2}$GeV) the ratio is much larger, about $10^{15}$. However, these numbers are highly model-dependent: in models beyond the standard model the number of degrees of freedom may be larger, but on the other hand, not all degrees of freedom in a particular theory necessarily interact with its defects---that number could even be zero.

It should be noticed that the two terms have generically different dependencies on the scale factor: the Hubble term behaves as $H\propto a^{-1/\lambda}$ (for a scale factor $a\propto t^\lambda$) while the friction term behaves as $\ell_f^{-1}\propto a^{-(1+n)}$. Therefore it follows that for a fast expansion or contraction rate, $\lambda(1+n)>1$ the Hubble term decays more slowly (and eventually dominates) in an expanding universe, and conversely it grows more slowly in a contracting universe. In the opposite regime of slow expansion or contraction, corresponding to $\lambda(1+n)<1$, it is the friction term that decays more slowly in the expanding case (and grows more slowly in a contracting one). Interestingly, the transition between these fast and slow regimes depends on the dimensionality of the defect: it occurs at $\lambda=1/2$ (that is, the radiation era) for monopoles, at $\lambda=1/3$ for cosmic strings, and at $\lambda=1/4$ for domain walls. This justifies our statements in the first paragraph of this section.

In any case, it is interesting to study the behavior of the defect networks in regimes where the friction term dominates. This has already been done, for radiation and matter-dominated expanding universes, both for cosmic strings \cite{MS1,MS2} and domain walls \cite{AWALL}. In that case the friction term is assumed to be dominating but decaying faster than the Hubble term, so the two possible solutions (known as the stretching and Kibble regimes) are both transient ones, and linear scaling eventually ensues. Here we provide a simple generalization of this analysis, first by considering all possible expansion rates (that is, whether $\lambda$ is fast or slow) and second by considering both expanding and contracting universes. The interesting result is that in all these cases not only do these two solutions still exist, but they are the only possible ones.

In this case, neglecting the Hubble damping term and keeping the friction one, the evolution equations become
\be
(4-n)\frac{dL_{inv}}{dt}=(4-n)HL_{inv}+\sigma\frac{v^2L_{inv}}{a^{1+n}} +cv \,
\ee
\be
\frac{dv}{dt}=(1-v^2)\left[\frac{k}{L_{inv}}-\sigma\frac{v}{a^{1+n}}\right]\,.
\ee
We have kept the invariant index in the lengthscale $L$ for completeness, although in this case the physical and invariant lengths are effectively the same: it is easy to show that the only possible attractor solutions of this system have decreasing non-relativistic speeds ($v\rightarrow0$), and therefore the Lorentz factor approaches unity. From this one can then find the two possible solutions. For low density, slow networks the energy loss (chopping term) is negligible and the network is simply conformally stretched (or contracted) as the universe evolves. This is therefore the stretching regime, in which we have
\be
L\propto a
\ee
\be
v\propto \frac{\ell_f}{L}\propto a^{n}\,.
\ee
Note that although the defect velocity is small, it is growing if the universe is expanding, and decreasing if the universe is collapsing. This implies that this solution must always be a transient one. If the chopping term cannot be neglected we have the Kibble regime \cite{Kibble1,addedC}
\be
L\propto \sqrt{ a^{1+n+\frac{1}{\lambda}}} \propto \sqrt{\frac{\ell_f}{|H|}}
\ee
\be
v\propto \frac{\ell_f}{L} \propto \sqrt{ a^{1+n-\frac{1}{\lambda}} } \propto \sqrt{\ell_f|H|}    \,.
\ee
In this case the correlation length is the geometric mean between the damping length and the horizon length. Note that for fast expansion rates $\lambda(1+n)>1$ the velocity increases in the standard case of an expanding universe (in which case the solution is a transient one, and is followed by the usual linear scaling one) but it would be an attractor for a contracting universe. Conversely, for slow expansion rates $\lambda(1+n)<1$ the velocity decreases in an expanding universe (in agreement with the fact that the friction term does dominate asymptotically in this case) whereas it increases in a collapsing universe, meaning this solution would be transient. 

\section{Conclusions}

In this work we introduced a new alternative formulation of the velocity-dependent one-scale model, based on a physical rather than invariant characteristic length scale. While the two formulations are equivalent (as we have explicitly shown) our discussion provides additional conceptual insight on the behavior of the model, and specifically on its phenomenological parameters.

As an application we studied one context in which such a formulation is particularly useful: the evolution of defect networks in contracting universes. We have shown that provided Hubble damping is the dominant damping term the networks become ultra-relativistic and are conformally contracted as the universe collapses (so that $L_{ph}\propto a$), and that asymptotically they will behave like a radiation fluid. We also carried out a general analysis of the friction-dominated case, showing that the stretching and Kibble regimes that have been identified for the case of expanding universes also apply to the contracting cases.

Physically, the interesting point to notice is that the evolution of a defect network in a contracting universe is not simply the reverse of that in an expanding one: instead, there is an unavoidable asymmetry, previously discussed in \cite{Cyclic}. Our results significantly generalize previous works and apply to extended defects (domain walls and cosmic strings), and with some caveats also to global monopoles. The scaling laws we have obtained can be tested in numerical simulations of sufficiently high resolution and dynamic range---an interesting task which we leave for subsequent work.

\begin{acknowledgments}

We are grateful to Ivan Rybak for helpful discussions on the subject of this work. This work was done in the context of project PTDC/FIS/111725/2009 (FCT, Portugal). CJM is also supported by an FCT Research Professorship, contract reference IF/00064/2012, funded by FCT/MCTES (Portugal) and POPH/FSE (EC).

\end{acknowledgments}

\bibliography{contracting}
\end{document}